\documentclass[a4paper,11pt]{article}
\usepackage{jcappub} 
\usepackage{txfonts}
\usepackage{mathtools}
\title{\boldmath Observational tests in scale invariance III: wide binary stars}

\author[1]{Andr\'e Maeder,}
\author[ 2, 3, 4]{Fr\'ed\'eric Courbin}
\affiliation[1]{Geneva Observatory, Chemin des Maillettes 51, CH-1290 Sauverny, Switzerland}
\affiliation[2]{Institute of Physics, Laboratory of Astrophysics, Ecole Polytechnique F\'ed\'erale de Lausanne (EPFL), Observatoire de Sauverny, 1290 Versoix, Switzerland}
\affiliation[3]{ICC-UB Institut de Ci\`encies del Cosmos, University of Barcelona,
Mart\'i Franqu\`es, 1, E-08028 Barcelona, Spain}
\affiliation[4]{ICREA, Pg. Llu\'is Companys 23, Barcelona, E-08010, Spain}

\emailAdd{Andre.Maeder@UniGe.ch}

\abstract{ Recent studies of wide binary stars based on Gaia DR3 suggest that the relative orbital velocities of objects with separations $s >$ 3'000 astronomical units are statistically larger than the standard Newtonian predictions. Obviously there is no Dark Matter halo arround binary stars that could be invoked to explain  these high velocities. However, we explore the properties of two-body systems in the framework of scale invariant vacuum theory, focusing on the case of objects with extreme separations. In this regime, the additional acceleration term present in the modified Newton equation with scale invariance becomes important, and may even dominate the dynamical evolution at very low gravities. Comparisons with Gaia DR3 observations of wide binaries are performed and suggest that binaries with  separations $s >$ 3'000 astronomical units have experienced such an evolution for a few Gyr, accounting well for the observed velocity excesses.}

\begin{document}
\maketitle
\flushbottom

\section{Introduction} 
\label{intro}

The scale invariant vacuum (SIV) theory rests on a fundamental and simple idea that extends the group of invariances subtending gravitation theory, as suggested by Dirac \cite{Dirac73}. In fact, General Relativity (GR) and Maxwell equations are indeed invariant in the empty space without charges and currents. The presence of matter in the Universe, by fixing some scales, tends to kill scale invariance. Cosmological models have been derived from
a general scale invariant equation confirmed by an action principle \cite{Canuto77,MaedGueor23}. They show that scale invariant effects fully disappear only above the mean critical density $\varrho_{\mathrm{c}}= 3 H^2_0 / 8 \pi G$ \cite{Maeder17a}. For typical Universe models with $\Omega_{\mathrm{m}}= 0.2$ or 0.3, the effects are very much reduced, being negligible in current life. However they do play a significant role in at least two cases: 1- very low density regions, and 2-  dynamical phenomena occruring at very long timescales.

In paper I of this series \cite{Maeder24},  the dynamical properties of galaxies in the scale invariant theory were studied. Among the many results,  it was found  that scale invariant dynamics lead to a large reduction of the mass estimates of galaxy clusters.  Account could be given to the relatively flat rotation curves of galaxies, to the steeper rotation curves observed in redshifted galaxies at $z=2$ \cite{Nestor Shachar23},   to the omnipresent relation between the apparent amount of dark and baryonic matters, and also to  effects in the "vertical" velocity dispersion in the Milky Way. Paper II  by Maeder and Courbin \cite{MCourbin24} was devoted to strong gravitational lensing. Using the exterior Schwarzschild solution, we derived the second order geodesic equations and showed that scale invariance  has no significant effects on gravitational lensing. However, we proceeded to a detailed studied of the system JWST-ER1 \cite{Dokkum23} and showed that, while the mass estimates by lensing are totally free of scale invariant effects (the geodesics are the same as in standard GR), it is not the case for the spectrophotometric mass determinations, which in standard theory are severely biased towards too low values. An analysis of the large Sloan Lens ACS (SLACS) sample of 100 galaxies  \cite{Auger09} was  further supporting this view.

In the present  study, we consider the case of the very wide binaries with separations $s \sim 10^4$ au, noting that such a separation corresponds to an orbital  period of 816'500 yr for a total binary mass of 1.5 M$_{\odot}$. 
In such wide-separation systems, the effective gravity nearly vanishes, becoming as low as in the outer regions of galaxies so that similar dynamical effects may possibly be expected. These questions have been boosted by the extraordinarily accurate results of the Gaia satellite, delivered first in the Gaia eDR3 and more recently by  Gaia DR3, so that since about a decade, a number of most interesting and detailed investigations have been performed. These are reviewed in Section~\ref{OBS} below. Section~\ref{orbital} presents the properties of orbital motions in  the scale invariant theory. Section~\ref{velocity} studies the loose and detaching systems. Section~\ref{comp} compares observations to the theoretical predictions. Finally, Section~\ref{concl} summarizes our conclusions.																									
\section{Current status of observational results on very wide  binaries}  
\label{OBS}

The relative independence of the orbital velocity from the orbital radius (e.g.  the relatively flat rotation curves of galaxies) has generally been interpreted in term of dark matter, while a minority of voices \cite{Milgrom83,Maeder17c} suggested possible gravity effects. It makes then sense to envisage that such effect should  be present in systems other than galaxies, and subject to very low gravitational fiels. We examine here some important recent studies on the dynamical properties of wide binary stars, where possible deviations from the Newtonian predictions have been investigated. As is often the case with studies of new physical effects,  at the limit of the experimental or observational possibilities, this is a domain that is very rich in different and sometimes controversial views.

Using a sample of 280 wide binaries observed by the Hipparchos satellite Hernandez \cite{Hernandez12} showed that relative velocities larger than Newtonian expectations were measured, independently of their separation. This result was supported by radial velocity observations  \cite{Scarpa17}. Pittordis and Sutherland \cite{Pittordis18} showed the possibility that future observations may solve the question and pointed out that some  MOND generalizations do not seem favourable. Banik and Kroupa \cite{BKroupa19} considered the case of Proxima Centauri around A and B Centauri and showed that a distinction between Newtonian and MOND predictions should become possible in the future. A  sample  of  929 carefully selected binary pairs  from the Gaia Early Data Release 3 eDR3 \cite{GaiaColl21} was studied with stringent quality requirements by Hernandez \cite{Hernandez22}, who found a Newtonian behaviour of the orbital velocities for gravities above about  $a_0 \geq 1.2 \times 10^{-10}$ m s$^{-2}$, while below this value the velocities remain constant, independently of the orbital radius. This value $a_0$ is the limit below which the so-called deep MOND limit, originally derived from galactic rotations, applies with a gravity given by $g = \sqrt{g_{\mathrm{N}} \, a_0}$, i.e. a stronger gravity than the Newtonian value $g_{\mathrm{N}}$ \cite{Milgrom83,Milgrom09}.

The case of undetected tertiaries is always a major concern in such studies. Clarke \cite{Clarke20}  considered that a fraction of 50\% of tertiaries could account for the results obtained by \cite{Hernandez19}. Indeed, many of the significant tertiaries could be eliminated \cite{Hernandez22} based on criteria drawn from simulations of multiple systems  by \cite{Belokurov20}. These authors showed that a RUWE value smaller than 1.4 (Renormalized Unit Weight Error, a value below 1.4 indicates a good astrometric solution), a high S/N-ratio together with HR diagram selection criteria maintain the ratio of tertiaries below 5\%. A still more extended and more accurate study of the internal kinematics of wide binaries was published by Hernandez \cite{Hernandez23} on the basis of the enhanced DR3 Gaia catalogue with 800'000 binary candidates. In the successive series of severe data quality cuts, Hernandez used the binary probabilty $B_p$ for each star based on spectral, photometric and astrometric information, the RUWE index, the radial velocities, the isolation criteria,  the Gaia FLAME work package for mass estimate, etc. Most stringent quality  criteria let  a set of 450 top-quality isolated binaries. On this basis, Hernandez \cite{Hernandez23} confirmed  his earlier results. 

Pittordis and Sutherland \cite{Pittordis23} examined the Newtonian and MOND predictions on the basis of a large sample of 73'159 wide binaries from the Gaia eDR3 objects within a distance of 300 pc and with a G magnitude brighter than 17. These authors  critically noticed that, although tight constraints on the sampling are performed by Hernandez et al., their surviving sample remained very small, offering only a poor statistical basis. With masses determined by a main-sequence mass-luminosity relation,  the analysis of  Pittordis and Sutherland is based on the histograms of the frequency  distributions as a function the ratios $\upsilon_p/\upsilon_c(r_p)$, where $\upsilon_p$ is the pairwise relative projected velocity and where $\upsilon_c(r_p)=\sqrt{GM_{tot}/r_p}$ is the estimated circular orbit velocity at the current projected separation $r_p$. The histograms showed a peak close to the Newtonian predictions, with in addition a long tail likely produced by unseen companions. The observed distributions were then fitted by simulated mixtures of binaries, triple and flyby systems with the Newtonian and MOND theoretical predictions. They noted, with some reserve, that standard gravity is preferred.

Chae \cite{Chae23} developed a Monte Carlo method to deproject the observed 2D  motions into the 3D relative space, in order to provide results in the acceleration plane. In addition, since it is inevitable that some binaries hide close inner companions, Chae proceeded to a calibration of the fraction $f_{\mathrm{multi}}$ of multiples within wide binaries. With $f_{\mathrm{multi}}$ taken as a free parameter,  the  determinations were iterated until the deprojected data at high acceleration, i.e. larger than $>10^{-10}$ m s$^{-2}$,	agreed with Newtonian expectations. A value $f_{\mathrm{multi}}$ in the range 0.3 to 0.5   was obtained, in agreement with current literature and then applied to the whole sample. The comparison of the kinematic acceleration $g= \upsilon^2/r$ and the Newtonian gravity $g_{\mathrm{N}}= G M_{\mathrm{tot}}/r^2$ showed  a ratio $g/	g_{\mathrm{N}}=1.43$ at $g_{\mathrm{N}}= 10^{-10.15}$ m s$^{-2}$, in agreement with AQUAL, one of the many generalisations of MOND. Regarding other works, Chae \cite{Chae23} remarked that  Hernandez \cite{Hernandez23} found deviations  from Newtonian predictions, which are increasing with separation $s$, and  would thus  be inconsistent with MOND. Chae \cite{Chae23} also pointed out that it is difficult to test a gravity theory with a sample displaying a narrow low-acceleration range as used by Pittordis and Sutherland \cite{Pittordis23}.

Tests of the MOND dynamics on the basis of wide binaries have also been performed by Banik, Pittordis, Sutherland et al. \cite{Banik24}. They showed that, locally, the orbital velocities predicted by this dynamics should exceed the Newtonian prediction by about 20\% at large separations (as the external galactic field limiting the effect). Using 8'611 wide binaries from Gaia DR3 within 250 pc of the Sun, selected according to a set of criteria on the consistency of the two parallax components, on the angular separations, on the projected velocity differences and on a number of other refined quality cuts, they proceeded to a detailed statistical analysis integrating the orbits, accounting for line-of-sight contamination and undetected close companions. According to these authors, the uncertainties of the relative velocity between stars have not been considered in the work by Chae \cite{Chae23}. They also  proceeded to a modeling of the orbits in the Newton and MOND approximations. Their statistical analysis based on frequency histograms in the line of the above mentioned work by Pittordis and Sutherland lead them to conclude to a strong preference for Newtonian gravity, although their best model does not fully reproduce the observations. They finally suggested that further modifications to MOND, introducing a new fundamental scale beyond $a_0$,  could possibly preserve  its galactic successes. We are personnally rather reluctant to accept further modifications of a theory each time there is an inconsistency with observations. In this context, we may remark that the scale invariant theory, which rests on a single and simple clear hypothesis, has no internal adjustment parameter, no ad'hoc tweaks. 

A new complementary study was made by Chae \cite{Chae24}, now considering the catalogue by El-Badry et al. \cite{El-Badry21} that provides a probability, $\Re$, of chance-alignment. Chae applied a very strong requirement that $\Re < 0.01$ corresponding to "pure binaries". He also applied cuts based on RUWE, proper motions, radial velocities, mass calibrations.  Chae pointed out that the estimate of multiple systems  $f_{\mathrm{multi}}$ by Banik et al. \cite{Banik24} was not calibrated on high acceleration binaries  and that these authors  found a value $f_{\mathrm{multi}}=0.70$ that was incompatible with the range currently obtained between 0.3 and 0.5 by many authors. He  also noticed that the analysis of \cite{Banik24} was compounded by the inclusion of chance-alignment cases. Testing the dynamical gravity $\upsilon^2/r$, Chae \cite{Chae24} obtained a  ratio $g/g_{\mathrm{N}}=1.49$ at $g_{\mathrm{N}}= 10^{-10}$ ms$^{-2}$. As a main result, the distributions of the sky-projected velocities as a function of the sky-projected separation $s$ for the so-called pure binaries were also presented.  The deviations from Newton's starts around $\log s=3.3$  (in au) increasing up to $\log s=3.8$, keeping then an apparently constant deviation by about 0.1 dex larger. This is shown in Fig.~\ref{Chae13}, which offers a very interesting possibility of comparison with theoretical expectations. The effects of changes in the selection parameters of the sample were examined, from which the authors concluded that their results were robust. 

The above observational situation is well representative of scientific developments at the frontier of knowledge, with uncertain and often contradictory conclusions and critical debates. It also reflects the usual trade-off that needs to be made between precision and accuracy: some scientists prefer large data samples with high statistical precision but loose quality cuts while others give preference to smaller samples with stringent quality requirements leading to high accuracy/reliability but with limited statistical power.  The above discussion is not any exception to the rule, but our goal is not to add to the debate. We rather focus on presenting a new interpretation of the data on wide binary stars in the framework of scale invariant theory, which contains no adjustable parameter and which holds without any need to invoke a plethora of ad'hoc variations and fixes when confronted to new observational tests. In fact, all the developments in the present work can be seen as theoretical predictions independent of data. It just happens that these predictions match very well the current data on wide binary stars and given their current precision and accuracy. 


\section{Properties of orbital motions in  the scale invariant theory}  
\label{orbital}


The basic assumptions of the scale invariant theory and the related demonstrations and equations have been presented in previous works of this series, in particular in Maeder \cite{Maeder17a}. We also recommend the study by Maeder and Gueorguiev \cite{MaedGueor23} where the fundamental equations for cosmology and weak fields have been demonstrated from an action principle. While the cosmological acceleration is important, the current effects in weak fields in the Newtonian approximation are generally negligible, except for very low densities and long evolutionary timescales.  The modified Newton equation of motion in these weak fields, that we use here as a starting point, is  \cite{MBouvier79,Maeder23},
\begin{equation}
 \frac {d^2 \bf{r}}{d \tau^2}  \, = \, - \frac{G \, M(\tau)}{ r^2} \, \frac{\bf{r}}{r}   + \frac{\psi(\tau)}{\tau_0}   \frac{d\bf{r}}{d\tau} \, \,.  
\label{Nvec4}
\end{equation}
It is expressed in the current time units (seconds, years). In this expression, $\tau_0$ is the age of the Universe, for which we take a value $\tau_0 = 13.8$ Gyr. There,  $\psi(\tau)$ is a numerical factor, 
\begin{equation}
\psi(\tau) \, = \, 
\frac{t_0-t_{\mathrm{in}}}{ [t_{\mathrm{in}} +   \frac{\tau}{\tau_0} (t_0-t_{\mathrm{in}})] }, \quad \quad  \mathrm{with} \; \; t_0=1,
 \; t_{\mathrm{in}}= \Omega^{1/3}_{\mathrm{m}}, \quad \mathrm{thus} \;\;
\psi_0=\psi(\tau_0) \, =1-\Omega^{1/3}_{\mathrm{m}} \,.
\label{phi}
\end{equation}
$\Omega_{\mathrm{m}}$ is the usual density parameter and $t$ is a time parameter appropriate for cosmological models (see \cite{Maeder17a}), varying between $t_{\mathrm{in}}$ at the Big-Bang and $t_0=1$ at  present. The  relation between $t$ and $\tau$  is defined by,
\begin{equation}
\frac{\tau - \tau_{\mathrm{in}}}{\tau_0 - \tau_{\mathrm{in}}} = \frac{t - t_{\mathrm{in}}}{t_0 - t_{\mathrm{in}}}\, ,
\end{equation}
expressing that the age fractions with respect to the present age are evidently the same in both timescales. This leads to
$t= [t_{\mathrm{in}} +   \frac{\tau}{\tau_0} (t_0-t_{\mathrm{in}})]$. The function $\psi(\tau)$ is dimensionless and varies between $(1-t_{\mathrm{in}})$ at the present time, $\tau_0$, and $(1-t_{\mathrm{in}})/t_{\mathrm{in}}$ at the origin, i.e. when $\tau=0$. We verify that the time derivative of $\psi(\tau)$ is,
\begin{equation}
\dot{\psi} =  - \frac{\psi^2}{\tau_0}.
\label{dpsi}
\end{equation}
In scale invariant theory, the gravitational potential $\Phi=GM/r$ appears as a more fundamental physical quantity than the mass, as the potential is scale invariant while the mass is not. As is the case in Special Relativity the mass may vary. The mass then behaves like,
\begin{equation}
M \rightarrow M_0 \times \frac{1}{\lambda},  \quad  \mathrm{with} \; \lambda = \frac{1}{t},  \quad \; \mathrm{giving} \; \; M(\tau)= M(\tau_0)\, [t_{\mathrm{in}}+\frac{\tau}{\tau_0}(1-t_{\mathrm{in}})], 
\label{mt}
\end{equation}
a property  already recognized by Dirac \cite{Dirac73,Dirac74} and Canuto et al. \cite{Canuto77}. However, while the mass may vary to infinity in Special Relativity, here the variations of the inertia are very limited, typically of the order of the percent over the last 400 Myr. Such a behaviour is also required by the action principle in the scale invariant context \cite{MaedGueor23}.

\subsection{Binet equation, angular mometum conservation and orbital motions} 
\label{Binet}

In previous works \cite{MBouvier79,Maeder23}, some properties of motion were examined in a system with the dimensionless time, $t$, used in cosmological equations and varying between $t_{\mathrm{in}} = \Omega^{1/3}_{\mathrm{m}}$ at the Big-Bang and the present time, $t_0$. For a cosmology with $\Omega_{\mathrm{m}}=0.30$, time $t$ would only vary between $t_{\mathrm{in}}=0.6694$ and $t_0=1$. While the cosmological equations are quite simple in this very constrained system of units, the usual time units $\tau$, in years, are more adapted for observational applications. In this case, time varies bewteen $\tau=0$ and $\tau_0 = 13.8$ Gyr.  Although the physics is exactly the same, this brings some changes in the formal writing of the  equations. Starting from the equation of motion  in currrent units, described in Eq.~(\ref{Nvec4}), we may write the corresponding two equations in plane polar  coordinates $(r,\vartheta)$, which leads to
\begin{eqnarray}
\ddot{r}-r\dot{\vartheta}^2  & = & -\frac{GM}{r^2}+\frac{\psi}{\tau_0} \frac{dr}{d\tau}, \\
r \ddot{\vartheta}+ 2 \dot{r} \dot{\vartheta} & = & \frac{\psi}{\tau_0} r \dot{\vartheta}.
\label{2pol}
\end{eqnarray}
When not specified, the quantity $\psi$ is $\psi(\tau)$ and  $M$ is $M(\tau)$. We easily verify that the 2$^{nd}$ of the above equations has for an integral,
\begin{equation}
\frac{\psi}{\tau_0} \,r^2 \dot{\vartheta} = \mathcal{L},
\label{consang}
\end{equation}

where $\mathcal{L}$ is a constant expressed in $[{\rm cm^2 \, rad \, s^{-2}}]$.  The above equation is a generalisation of the classical angular momentum conservation $r^2 \dot{\vartheta} = {\rm const.}$ It shows that the classical angular momentum, $r^2 \dot{\vartheta}$, slowly increases with time following, 
\begin{equation}
r^2 \dot{\vartheta} = \frac{\tau_0 \,\mathcal{L}}{1-t_{\mathrm{in}}} \, [t_{\mathrm{in}}+\frac{\tau}{\tau_0}(1-t_{\mathrm{in}})],
\label{consangvar}
\end{equation}
a behaviour quite consistent with the evolution of radius and orbital velocity (see Sect.~\ref{vel}).
Our aim now is to express the contents of the three above equations into one single equation expressing the orbital motion of a test particle near a massive body 
of mass $M$.  Using Eq.(\ref{consang}) the derivative of $r$ with respect to $\vartheta$ can be written,
\begin{equation}
r' \equiv \frac{dr}{d \vartheta} = \frac{dr}{d\tau} \, \frac{d\tau}{d \vartheta}= \frac{\dot{r}\; r^2 \psi}{\mathcal{L} \; \tau_0} \quad
\mathrm{thus} \; \; \dot{r}=\frac{r' \mathcal{L} \, \tau_0}{r^2 \,   \psi}\,.
\label{dp}
\end{equation}
This kind of relation between $r'$ and $\dot{r}$ can be extended to $\ddot{r}$,
\begin{equation}
\ddot{r}= \frac{\mathcal{L}^2 \,r'' \tau^2_0}{r^4 \, \psi^2} - \frac{2 r'^2 \mathcal{L}^2 \tau^2_0}{r^5 \psi^2}+
\frac{r' \mathcal{L}}{r^2}\,,
\label{dd}
\end{equation}
thanks to (\ref{dpsi}). Now, we express the first equation  of (\ref{2pol}) with this last expression and get,
\begin{equation}
\frac{\mathcal{L}^2 \, \tau^2_0}{r^4 \, \psi^2}\left(r''-2 \frac{r'^2}{r}\right) -\frac{\mathcal{L}^2 \, \tau^2_0}{r^3 \, \psi^2}+
 \frac{G \, M}{r^2}=0 \,,
\label{r2}
\end{equation}
where remarkably and fortunately  the last  terms in Eqs.~(\ref{Nvec4}) and (\ref{dd}) have simplified. Now, writing this last equation with $u=1/r$ is leading to,
\begin{equation}
u'' + u = \frac{GM\, \psi^2}{\mathcal{L}^2 \, \tau^2_0}\,.
\label{Binet}
\end{equation}
This is the modified form of the well-known Binet equation, describing planetary trajectories. The classical form does not contain the term $\psi^2/ \tau^2_0$. If the 2$^{nd}$ member is equal to $0$, the solution is $u= C \cos \vartheta$, the polar equation of a straight line. Now, we search a solution of the form $u= p+C \cos\vartheta$ and get 
\begin{equation}
p=  \frac{GM\, \psi^2}{\mathcal{L}^2 \, \tau^2_0},
\end{equation} 
expressed in $[{\rm cm}^{-1}]$. This quantity, $p$, is often called the {\emph{latus rectum}} in  the  geometrical construction of a conic. $C$ is also a constant in $[{\rm cm}^{-1}]$. Thus, the solution $r(\vartheta)$ is
\begin{equation}
r=\frac{1}{\frac{GM\, \psi^2}{\mathcal{L}^2 \, \tau^2_0}+C \cos \vartheta}\, \quad \mathrm{or} \; \;
r= \frac{r_0}{1+e \cos \vartheta}  , \quad \quad \mathrm{with}  \; \; e= \frac{\,C \mathcal{L}^2 \, \tau^2_0}{GM\, \psi^2}\,,
\label{traj}
\end{equation}
where $e$ is the eccentricity essentially determined by the initial conditions. It is dimensionless and is therefore a scale invariant quantity. We see that if the eccentricity is zero, then $r_0=1/p$, 
\begin{equation}
r_0  \, = \, \frac{\mathcal{L}^2 \, \tau^2_0}{GM\, \psi^2}\,,
\label{r0}
\end{equation}
$r_0$ being  the radius  of the  circular orbit obtained when $e=0$. We verify that $r_0$ is expressed in $[{\rm cm}]$ and  behaves like $1/\lambda$, implying that  it is increasing with time (see next subsection).  For elliptical solutions, with $e<1$, the semi-major and semi-minor axes respectively are $a=\frac{r_0}{(1-e^2)}$ and $b=\frac{r_0}{\sqrt{1-e^2}}$, which behave in time like $r_0$. Thus, the solutions are not classical conics with constant parameters (except for $e$). Instead, as the result of the time dependence of some quantities like $r_0$, they  are {\emph{expanding conics}} as originally demonstrated by \cite{MBouvier79}.
 
\subsection{Secular properties of  the expanding conics and Third Kepler's Law}

Let us examine the time evolution of $r_0$, which determines the properties of the elliptical orbits, noting that similar relations would apply for parabolic and hyperbolic motions. From (\ref{r0}), one has  for the relatives variations of the  semi-major axis $a$,
\begin{equation} 
\frac{\dot{a}}{a}= -\frac{\dot{M}}{M}-2 \frac{\dot{\psi}}{\psi}\,.
\label{ad}
\end{equation}
From Eq.~(\ref{mt}), we have
\begin{equation}
\dot{M}=  \frac{M(\tau_0)}{\tau_0} (1- t_{\mathrm{in}}), \quad \quad \mathrm{and} \; \; \frac{\dot{M}}{M}= \frac{\psi}{\tau_0}\, ,
\end{equation}
so that with (\ref{dpsi})
\begin{equation} 
\frac{\dot{a}}{a}= -\frac{\psi}{\tau_0}+2 \frac{\psi}{\tau_0}=  \frac{\psi}{\tau_0} \,.
\label{aad}
\end{equation}
This implies that the semi-major axis $a(\tau)$ of the elliptical motions behaves like,
\begin{equation}
a(\tau)= a(\tau_0)\,  [t_{\mathrm{in}}+\frac{\tau}{\tau_0}(1-t_{\mathrm{in}})].
\label{atau}
\end{equation}
The semi-major axis is therefore increasing linearly in time $\tau$, in a way which is perfectly consistent with the variation of $a(t)=a(t_0) \times t $ previously obtained in the $t$-scale \cite{MBouvier79}. Due to the relations recalled above between the dimensional parameters $a$, $b$ and $r_0$, the above dependence on time $\tau$ also applies to all dimensional quantities.

We now turn to the orbital period $T$, which for simplicity is considered in the circular case. With the angular velocity
$\dot{\vartheta}= 2 \pi/T$, (for an elliptical orbit $\dot{\vartheta}$ would be the mean angular velocity)  and 
the angular momentum conservation $\dot{\vartheta}=\frac{\mathcal{L} \tau_0}{r^2 \psi}$, one is led to
\begin{equation}
\frac{\dot{\vartheta}}{\vartheta}= -2 \frac{\dot{r}}{r} -\frac{\dot{\psi}}{\psi}=
-2  \frac{\psi}{\tau_0} +  \frac{\psi}{\tau_0}= - \frac{\psi}{\tau_0},\quad \quad  \quad \mathrm{thus} \; \;\frac{\dot{T}}{T}= 
-\frac{\dot{\vartheta}}{\vartheta}= \frac{\psi}{\tau_0}.
\label{TT}
\end{equation}
This implies that the period $T$ is varying like,
\begin{equation}
T(\tau)= T(\tau_0)\,  [t_{\mathrm{in}}+\frac{\tau}{\tau_0}(1-t_{\mathrm{in}})]\,,
\label{Ttau}
\end{equation}
also increasing linearly in time $\tau$. Thus, we may note that the mass $M$, the dimensional  parameters $a$, $b$, $r_0$ and the orbital period $T$ all have the same kind of time dependence.

We may wonder what is happening to the Third Kepler's Law. From the conservation law (\ref{consang}) and for simplicity  still considering  a test particle with circular orbits with radius given by (\ref{r0}) we have,
\begin{eqnarray}
\mathcal{L}=\frac{\psi}{\tau_0} \,r^2_0 \frac{2 \pi }{T},  \quad \mathrm{and} \; \; r_0 = \frac{\mathcal{L}^2 \, \tau^2_0}{GM\, \psi^2} & = &
r^4_0 \frac{4 \pi^2}{T^2 \, GM}, \\
\mathrm{we \; get}   \quad \quad   \frac{r^3_0}{T^2}  &=& \frac{ G M}{4 \pi^2}\,,
\end{eqnarray}
which corresponds to the Third Kepler's Law for the case considered. This applies at any time $\tau$ since all the dependences in the form of the above $T(\tau)$ cancel each other. Remarkably, the Third Kepler's Law which originally has led to the Newton Law is still valid in the scale invariant theory, even though the Newton Law no longer applies as such. Thus, the Third Kepler Law is a most fundamental property of gravitation, a Rock in the theory of gravitation, independent of whether gravitation is scale invariant or not. This is the law which allows the determination of planetary masses in the Solar System and it remains unmodified by scale invariance, providing the correct masses at present time.
	
\section{The orbital velocities in bound and loose wide binaries}  
\label{velocity}

In relation with the velocity observations in wide binaries reviewed above, we now specify the behaviour of the orbital velocities in  a bound system consisting of a test particle around a central mass $M$. For masses $M_1$ and $M_2$, the sum $M_{\mathrm{tot}}$ of the two masses has to be accounted for. Again we consider the case of circular orbits. We have to distinguish the case of bound systems where Newtonian gravity $GM/r^2$ dominates over the so-called dynamical gravity $\frac{\psi}{\tau_0} \upsilon$ and the opposite case of loose systems governed by the dynamical gravity.

\subsection{The orbital velocities in a  bound two-body system}  
\label{vel}

Starting from the conservation of the angular momentum  $ \frac{\psi}{\tau_0} \, r^2 \dot{\vartheta} = \mathcal{L} \,$ according to (\ref{consang}), we have for the orbital velocity $\upsilon=r\dot{\vartheta}= \frac{\mathcal{L} \,\tau_0}{\psi \, r}$. Again considering a circular orbit, by introducing the expression of $r_0$ given in Eq.~(\ref{traj}) one obtains,
\begin{equation}
\upsilon = r_0 \dot{\vartheta} = \frac{\mathcal{L} \,\tau_0 \,GM \psi^2 }{\psi \,  \mathcal{L}^2 \,\tau^2_0 }= \frac{ \ GM \psi}{  \mathcal{L} \tau_0 }.
\label{vps}
\end{equation}
Expressing now the time dependence  of both $M$ and $\psi$, we simplify the above expression,
\begin{equation}\
\upsilon = \frac{G \, M(\tau_0)}{\mathcal{L}} \left(\frac{1-t_{\mathrm{in}}}{\tau_0}\right) = const.
\label{ups}
\end{equation}
The time dependences vanish, thus the orbital velocity is a scale invariant quantity, which is keeping constant in time during  secular orbital expansion. A more usual and convenient form may be obtained from (\ref{vps}) and (\ref{traj}),
\begin{equation}
\upsilon^2=  \frac{ \ G^2 M^2 \psi ^2}{  \mathcal{L}^2 \tau^2_0 }= \frac{GM}{r_0}.
\label{u2}
\end{equation}
Both $M$ and $r_0$ have the same secular variations  defined by the evolution of  the scale factor, and both vary with $[t_{\mathrm{in}}+\frac{\tau}{\tau_0}(1-t_{\mathrm{in}})]$. Thus, the secular evolution of the scale factor $\lambda$ has no incidence on the velocity which remains constant, as already pointed out. Moreover, the resulting velocity in the scale invariant two-body system has  the same expression as the classical one. It is to be emphasized that the fact that the instantanous  expressions of the velocity at time $\tau$ are the same as the standard ones does not imply that the velocity dispersion $\overline{\upsilon^2}$ of gravitational systems is the same in the classical and SIV theory. 

In standard theory, a body drifting away would see its velocity decrease  according to the standard law of angular momentum conservation $r_0 \times  \upsilon = const. $ Here in the scale invariant theory, the angular momentum conservation law is $\frac{\psi}{\tau_0} \, r_0 \times \upsilon= const.$  according to (\ref{consang}). For a rapid change of the radius (rapid means in a negligible time in front of $\tau_0$), the effect would evidently be the same as in the standard case. On longer timescales, the secular changes of  the radius do leave the velocity constant, so that the angular momentum  $ r_0 \times \upsilon$ just varies like $r_0$, quite consistently with (\ref{consang}) and the comments following this  conservation law.

In this section, we have seen (in the current units) the main properties of the two-body system where both the Newtonian  $\frac{GM}{r^2}$ and the dynamical $\frac{\psi(\tau)}{\tau_0} \times \upsilon$ gravities  are significant. The extreme case of very wide binaries, where the dynamical acceleration may largely dominate over gravity, is studied below.

\subsection{Loose and detached systems}  
\label{lloo}

The gravity field of a mass $M$ extends to infinity. For increasing distances from $M$, the attraction of other masses in the Galaxy makes the local gravity to vanish. This occurs at  the Jacobi radius, also called Roche radius, where the Galactic field starts dominating; for a typical system a value of about 1.70 pc \cite{Jiang10} is estimated for this radius. Long before this  happens, the dynamical acceleration  $\frac{\psi}{\tau_0} \times \upsilon $  may  overcome the  ambiant Newtonian gravity  $g_{\mathrm{N}}$. The equality of the two terms of the equation of motion (\ref{Nvec4}) occurs for a gravity  $g_0$, as given by  \cite{Maeder23}
\begin{eqnarray} 
 g_0 \, = \, 
 \frac{(1-\Omega_{\mathrm{m}})^2}{4} \, n\,  c \, H_0\,  \quad \quad 
 \mathrm{or} \quad  g_0 \,= \, \frac{n\, c \, (1-\Omega_{\mathrm{m}}) (1-\Omega^{1/3}_{\mathrm{m}})}{2\, \tau_0}\, .
 \label {go}
 \end{eqnarray}
The product $c \times H_0$ is equal to $6.80 \times 10^{-8}$ cm s$^{-2}$. For $\Omega_{\mathrm{m}}$=0, 0.10, 0.20, 0.30 and 0.50, we get $g_0 =$ (1.70, 1.36, 1.09, 0.83, 0.43) $ \times \, 10^{-8}$ cm s$^{-2}$ respectively. Interestingly enough, this value of $g_0$  well corresponds  to the value of $a_0$ of about 1.2 $\times \,10^{-8}$ cm s$^{-2}$  characterizing the deep MOND  limit \cite {Milgrom83,Milgrom09}, where the gravity is $g= \sqrt{a _0 \; g_{\mathrm{N}}}$, where $g_{\mathrm{N}}$ is the Newtonian gravity. For gravities below $g_0$, the equation of motion (\ref{Nvec4}) acccepts, {\emph{as an approximation}}, a similar relation of the form $g= \sqrt{g _0 \; g_{\mathrm{N}}}$ \cite{Maeder23}, with $g_0$ given  by  (\ref{go}). This approximation is valid   over the last  400 Myr with an accuracy 1-2\% depending on $\Omega_{\mathrm{m}}$. For a given total mass $M_{\mathrm{tot}}$ of a binary system, the equality of the Newtonian and dynamical accelerations occurs at a separation,
\begin{equation}
s_{\mathrm{trans}}\, = \, \sqrt{\frac{G \,M_{\mathrm{tot}}}{g_0}} \,. 
\label{strans}
\end{equation}
For M$_{\mathrm{tot}}= 1.0$  M$_{\odot}$, we get a value of $s_{\mathrm{trans}}= 7.0$ [kau] or 0.034 [pc] (8.6 [kau] or 0.042 [pc] for a typical value of the total mass of 1.5 M$_{\odot}$) \cite{Chae24}.  

What happens for wide binaries  approaching  the transition? We may qualitatively understand the  effects of a declining   Newtonian gravity by considering the effects of a decrease of the product $G \times M$ in the basic parameters  $e$ and $r_0$. From Eqs. (\ref{traj}) and (\ref{r0}), we see that  both  the eccentricity $e$ and the radius of the circular orbit (or  latus-rectum of an ellipse) $r_0 \, \rightarrow \, \infty$ for a product $G \times M$ approaching zero. This means that  the orbit first extends in size together  with an increase of the ellipticity up to $e=1$, where it becomes parabolic. At the parabolic limit, we have,
\begin{equation}
C \, = \,  \frac{GM\, \psi^2}{\mathcal{L}^2 \, \tau^2_0}      \, \equiv   \,\frac{1}{r_0}\,.
\end{equation}
Then, for a further decline of Newtonian gravity, the eccentricity  becomes larger than 1 and  the orbit is hyperbolic tending to a straight line. It is interesting to examine the behaviour of the orbital velocity in this limit. Within SIV theory, it is evidently preferable  to use the full  equation of motion and the appropriate relations  rather than the MOND-like approximation. From~(\ref{Nvec4}), we have,
\begin{equation}
 \frac {d^2 \bf{r}}{d \tau^2}  \, =  \frac{\psi(\tau)}{\tau_0}   \frac{d\bf{r}}{d\tau} \,,    \quad \quad \mathrm{or} \; \; 
 \frac{d \upsilon}{d\tau}  \,= \,   \frac{\psi(\tau)}{\tau_0} \; \upsilon \,,
\label{Nvec5}
\end{equation}
for the linear motion. This can also be rewritten as,
\begin{equation}
\frac{d\upsilon}{\upsilon}  \,= \,   \frac{\psi(\tau)}{\tau_0} \; d\tau\,.
\label{Nvec6}
\end{equation}
From Eq.~(\ref{dpsi}), one has
\begin{equation}
\psi \, = \, - \frac{\dot{\psi}}{\psi} \tau_0\,, \quad  \mathrm{and \; thus}  \; \;
\frac{d\upsilon}{\upsilon}  \,= \, -  \frac{\dot{\psi}}{\psi} \, d\tau\, = \, -  \frac{d\psi}{\psi}\,.
\end{equation}
The integration gives according to (\ref{phi}),
\begin{equation}
\ln \upsilon = - \ln \psi + const.  \quad \quad \mathrm{thus }   \; \; \upsilon =  \frac{k}{\psi}= 
\frac{k}{(1-t_{\mathrm{in}})}\, [t_{\mathrm{in}}+\frac{\tau}{\tau_0}(1-t_{\mathrm{in}})]\,,
\end{equation}
where $k$ is a constant with the dimension of a velocity. Setting $\tau=\tau_0$, we have $\upsilon(\tau_0)=\frac{k}{(1-t_{\mathrm{in}})} $ and the velocity becomes,
\begin{equation}
\upsilon(\tau) = \upsilon(\tau_0)\,  [t_{\mathrm{in}}+\frac{\tau}{\tau_0}(1-t_{\mathrm{in}})].
\label{vt}
\end{equation}
Thus, we see that the velocity is no longer keeping constant with secular variations as in the bound system, but now the velocity is increasing linearly in time $\tau$ under the effects of the dynamical acceleration $ \frac{\psi(\tau)}{\tau_0} \; \upsilon$. Turning now to the covered distance $r$ during this type of accelerated motion, we have from (\ref{vt}),
\begin{equation}
\frac{dr}{d\tau} = \upsilon(\tau_0)\,  [t_{\mathrm{in}}+\frac{\tau}{\tau_0}(1-t_{\mathrm{in}})], \quad \mathrm{thus} \; \;
r(\tau) = \frac{1}{2} \frac{\tau_0}{(1-t_{\mathrm{in}})}\, \upsilon(\tau_0)\, [t_{\mathrm{in}}+\frac{\tau}{\tau_0}(1-t_{\mathrm{in}})]^2.
\label{rt}
\end{equation}

 This is quite a classical type of accelerated motion, increasing with the square of the time.
Thus, when the dynamical gravity $\frac{\psi}{\tau_0}  \upsilon$ significantly dominates over the overall Newtonian gravity, 
the motion becomes hyperbolic, with an eccentricity $e>1$, and a velocity increasing  linearly with time. As a result, the  covered distance fastly increases with a quadratic dependence in time,
\begin{equation}
\delta(\tau) = [t_{\mathrm{in}}+\frac{\tau}{\tau_0}(1-t_{\mathrm{in}})]^2.
\end{equation}
This expression in $t^2$ is shared by several equations in scale invariance, giving the velocity, the orbital radius and the mass, as also seen in Eq.~(\ref{mt}). Let us call it the \emph{dynamical regime}. In order to numerically estimate the possible changes in velocities and in covered distances in this regime,  let us consider a reference time $\tau_{\mathrm{N}}$  schematically marking the end of the Newtonian  regime and the beginning of the dynamical one. We then get the following ratios,
\begin{equation}
\frac{\upsilon(\tau)}{\upsilon(\tau_{\mathrm{N}})}= \frac{[t_{\mathrm{in}}+\frac{\tau}{\tau_0}(1-t_{\mathrm{in}})]}
{ [t_{\mathrm{in}}+\frac{\tau_{\mathrm{N}}}{\tau_0}(1-t_{\mathrm{in}})]}, \quad \quad \mathrm{and}\; \; 
\frac{r(\tau)}{r(\tau_{\mathrm{N}})}= \frac{[t_{\mathrm{in}}+\frac{\tau}{\tau_0}(1-t_{\mathrm{in}})]^2}
{ [t_{\mathrm{in}}+\frac{\tau_{\mathrm{N}}}{\tau_0}(1-t_{\mathrm{in}})]^2},
\end{equation}
and thus 
\begin{equation}
\frac{\upsilon(\tau_0)}{\upsilon(\tau_{\mathrm{N}})}= \frac{1}{ [t_{\mathrm{in}}+\frac{\tau_{\mathrm{N}}}{\tau_0}(1-t_{\mathrm{in}})]},
\quad \quad \mathrm{and}\; \; 
\frac{r(\tau_0)}{r(\tau_{\mathrm{N}})}= \frac{1}{ [t_{\mathrm{in}}+\frac{\tau_{\mathrm{N}}}{\tau_0}(1-t_{\mathrm{in}})]^2}.
\label{vn}
\end{equation}
This gives the change in velocity and radius, between the transition time  $\tau_{\mathrm{N}}$ to today in the dynamical regime. Of course, binary systems are observed locally in the Milky Way at the present time, i.e. at time $\tau_0$.


\section{Comparisons of observations and theory for very wide binaries} 
\label{comp}

We now  proceed to numerical estimates of the above theoretical predictions and we compare them with the observations by Chae \cite{Chae24} and by Hernandez \cite{Hernandez23}. Both works consider a vast range of binary stars samples in Gaia DR3 and show deviations  from Newtonian predictions in the observed relation between velocities  $\upsilon$ and separations $s$.

\begin{figure*}[t!]
\centering
\includegraphics[width=.70\textwidth]{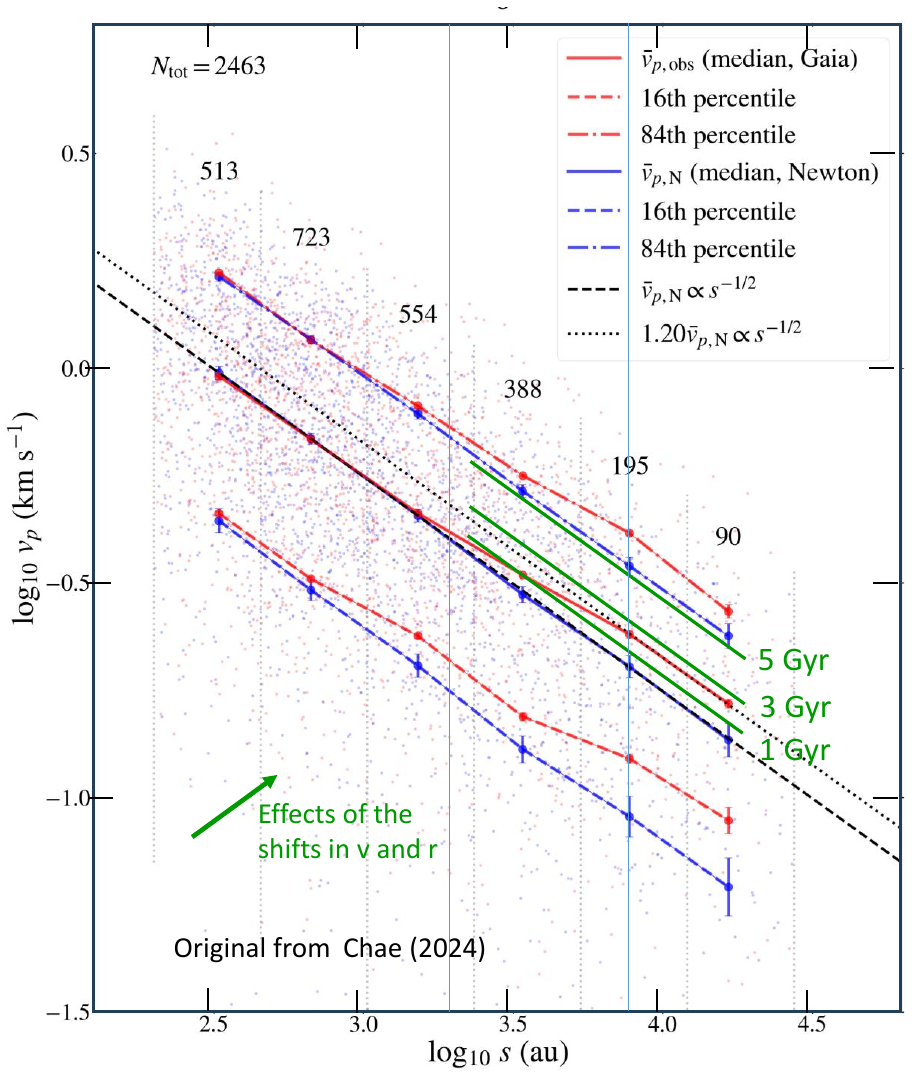}
\caption{Projected velocities $\upsilon_p$ as a function of separation $s$ for the main sample in Fig.~13 from Chae \cite{Chae24}. The very small red dots are the observed values and the blue dots are the Newtonian values in the Monte-Carlo analysis. The larger dots and their connecting lines are the median and percentiles as indicated. The central black dashed line shows the Keplerian relation in $1/\sqrt{s} $. The green lines are additional indications of the mean deviations along the hyperbolic path of the loose systems after 1, 3 and 5 Gyr from the time, $\tau_N$, where they depart from the Newtonian relation evolving under the dynamical acceleration only. The direction of the effects in velocity and separation is indicated by a green arrow at the bottom on the left side. The deviation of a given system increases linearly with time as indicated by Eqs~.(\ref{vn}). We see that the mean observed relation for the loose systems correponds to a mean  evolution during about  2 to 3 Gyr. In other words, in only a few Gyrs, loose systems drift away from the Newtonian relation in a way fully compatible with the dynamical evolution described by scale invariance (Figure adapted from Chae \cite{Chae24}).}
\label{Chae13}
\end{figure*}

\subsection{Comparison with the analysis by Chae \cite{Chae24}}

This analysis by Chae \cite{Chae24} has been presented in Sect.~\ref{OBS}. Fig.~\ref{Chae13} shows  the logarithmic relation between $\log \upsilon$ and $\log s$ for a sample of 2'463 binary pairs with masses estimated from relations between the absolute G-band magnitude, $M_G$, and the mass, $M$. The mean of the observations at different  separations, $s$, are given by the continuous red lines. The blue lines result fom Monte-Carlo (MC) simulations. In practice, from an observed  set of masses $M$ and $s$ values, each MC realization draws the velocity from the classical Newtonian equations with the appropriate statistical distributions of the orbital parameters, namely  $\vartheta$ the longitude of the observed star, $\vartheta_0$ the longitude of the periastron, the orbit inclination $i$ and $e$ the eccentricity. A serie of 400 MC simulations are performed to estimate the statistical error bars, as indicated on the blue line.

The average total mass of the binary systems is around 1.5 M$_{\odot}$, meaning that the Main-Sequence lifetimes are generally several Gyr. Let us consider 3 cases: when the systems start deviating from Newton's Law  1, 3 and 5 Gyr ago. This corresponds to 3 values of the past time $\tau_{\mathrm{N}}=$ 12.8,  10.8 and  8.8  Gyr. These ages  lead to the following velocity ratios respectively $\frac{\upsilon(\tau_0)}{\upsilon(\tau_{\mathrm{N}})}= 1.048, 1.159$ and 1.296. Thus, according to Eqs.~(\ref{vn}), the shifts  in velocities and distances since time, $\tau_{\mathrm{N}}$, when the systems  entered the dominant free regime, are the following ones,
\begin{equation}
\log \left[\frac{\upsilon(\tau_0)}{\upsilon(\tau_{\mathrm{N}})}\right]=0.020, 0.064, 0.113, \quad \quad  \mathrm{and} \quad
\log \left[\frac{r(\tau_0)}{r(\tau_{\mathrm{N}})}\right]=0.040, 0.128, 0.226.
\label{val}
\end{equation}
\begin{figure*}[t!]
\centering
\includegraphics[width=0.70\textwidth]{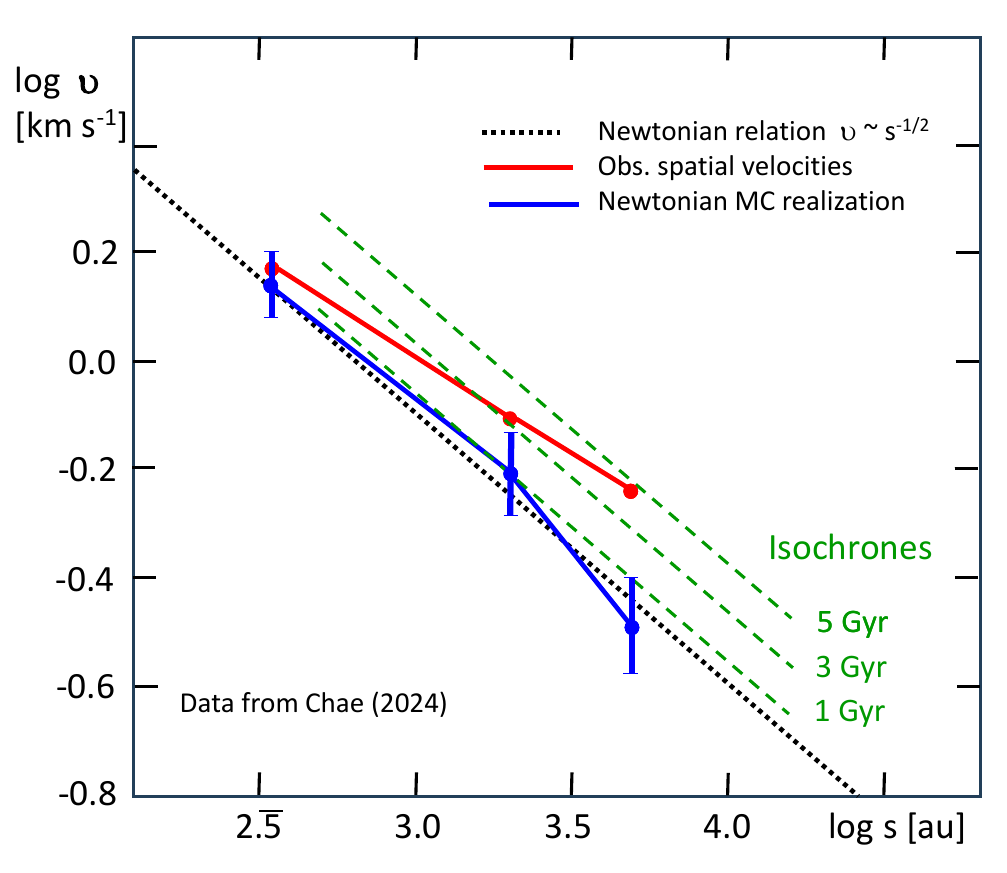}
\caption{Comparisons of observations and theory for a sample of 40 very wide binaries with exceptionally precise radial velocities, with individual relative errors smaller than 0.005, as selected by Chae \cite{Chae24}.  The green broken lines shows the isochrones corresponding to departures from the Newtonian law after 
 1, 3 and 5 Gyr of evolution under the dynamical acceleration in the scale invariant theory. The departure from Newton's relation is progressive and tends towards a value between 3 and 5 Gyr (Data are from Chae \cite{Chae24}).}
\label{Chae21}
\end{figure*}

In Fig.~\ref{Chae13} , the above shifts in $\log \upsilon$ and $\log r$ produce a displacement in the direction indicated by the green arrow in the lower left corner. The amplitudes of the displacements are determined by the above numerical values  of velocities and distances.  We see that the observed mean shift progressively increases between $s= 2$ [kau]  to 8  [kau], an interval which well encompasses the value of $s_{\mathrm{trans}}$ discussed in Sect.~\ref{lloo}, i.e. $log(s_{\mathrm{trans}}) = 3.84$ for 1 M$_{\odot}$ and $log(s_{\mathrm{trans}}) = 3.9$ for 1.5 M$_{\odot}$. Binary systems with ages corresponding to the green time-lines/isochrones may arise from binaries at different locations on the reference line $\upsilon \sim 1/\sqrt{s}$, the amplitudes of the shifts are expected to be  significant  somewhere above the limit $s_{\mathrm{trans}}$, as is observed  for separations larger than 10'000 au. Note that according to the Third Kepler's Law, a system with a total mass of 1.5 M$_{\odot}$ and a separation 10'000 au,  if still bound, would have an orbital period $T= 8.165 \times10^5$ yr.

The observed mean shifts corresponds to  an evolution in the free regime by about 2 to 3 Gyr. These shifts are clearly larger than the error bars as illustrated in the figure. However, we must be careful not to overinterpret these results because the velocities for separations $s$ larger than $10^4$ au  are of the order of 170 m/s  and the shifts in velocities correspond to velocity differences of about 27 m/s. Nevertheless, the shifts appear significant. This is particularly true since all the possible variations/combinations of the selection criteria by \cite{Chae24} all give similar shifts between the reference line  $\upsilon \sim 1/\sqrt{s}$ and the observed values near and above $s \sim 10^{3.5}$ au. As a reminder, these selection criteria include J-band photometry instead of G-band, different magnitude and mass domains, the use of more stringent astrometric requirements, the change of eccentricity distributions, and a range of estimated frequency of multiple systems. We may remark that even the initial general sample of 26'615 binary pairs shows, according to Chae's results, some deviation from the Newtonian behaviour for $s$-values larger than $\log s  \geq 3.5$.

Interestingly enough, after considering the statistical properties of  large samples as in Fig.~\ref{Chae13}, 
Chae \cite{Chae24} also analysed a smaller sample that includes only the 40 most highly accurate radial velocity data, with a relative precision equal or better than  0.005 on the radial velocities, i.e., comparable to the astrometric data. Three bins are formed from these exceptional 40 systems.  The statistical uncertainties are evidently larger, as shown on the blue line, but thanks to the exceptional data quality, the highest $s$-bin at $\log s = 3.7$ now shows a large deviation from Newtonian relation, which is well compatible with a free motion during nearly 5 Gyr. This unique highly accurate sample is reinforcing the conclusion that the existence of significant deviations from Newtonian relation for $\log s \geq 3.5$ and above.  The direction and size of the observed effects are compatible with a dynamical  evolution for several Gyr, dominated by the scale invariance term of Eq.~(\ref{Nvec6}).  At this stage, whether the mean duration of this phase is actually 3 Gyr, 5 Gyr or more remains uncertain. In any case, it may correspond to a substantial fraction of the life of our Galaxy.

To summarize, the full sample of 26'615 binary pairs, the sub-sample of 2'463 binaries with excellent data and the very restricted sample of 40 binary pairs with an extremely high precision of 0.005 on radial velocities all confirm the occurrence of significant deviations to Newton's predictions in the velocity vs.  $s$ relations, for separations larger than a few kau. These deviations correspond to an evolution in the dynamical regime for several Gyr and remain robust against any change made on statistical parameters when selecting sub-samples of pairs.

\subsection{Comparison with the analysis by Hernandez \cite{Hernandez23}}

Let us now turn to the sample of wide binaries assembled by Hernandez \cite{Hernandez23}, which results from very stringent quality criteria briefly recalled in Sect.~\ref{OBS}. In most  studies, the mass of the binary components was estimated from simple magnitude-mass scaling relations. More recently, mass estimates became available from an internal Gaia FLAME work-package making use of spectroscopic information in DR3. This provides more accurate masses for 24\% of the binary components. Comparisons of the masses obtained from magnitude-mass scaling with FLAME masses shows a small offset of about 0.05 M$_{\odot}$ below 0.7 M$_{\odot}$, which is then corrected by the author. The relative velocity $\Delta V$ analysis  restricted to the plane of the sky is primarily determined  through Gaia \emph{proper motions in two perpendicular directions}, radial velocities playing  a secondary role for separations $s < 0.1$ pc.  Final cuts are made based on  the signal-to-noise ratios $\Delta  V/\sigma_{\Delta V}$, yielding a high-quality sample of 450 highly-isolated wide binaries, which serve for the astrophysical study.

\begin{figure*}[t!]
\centering
\includegraphics[width=0.8\textwidth]{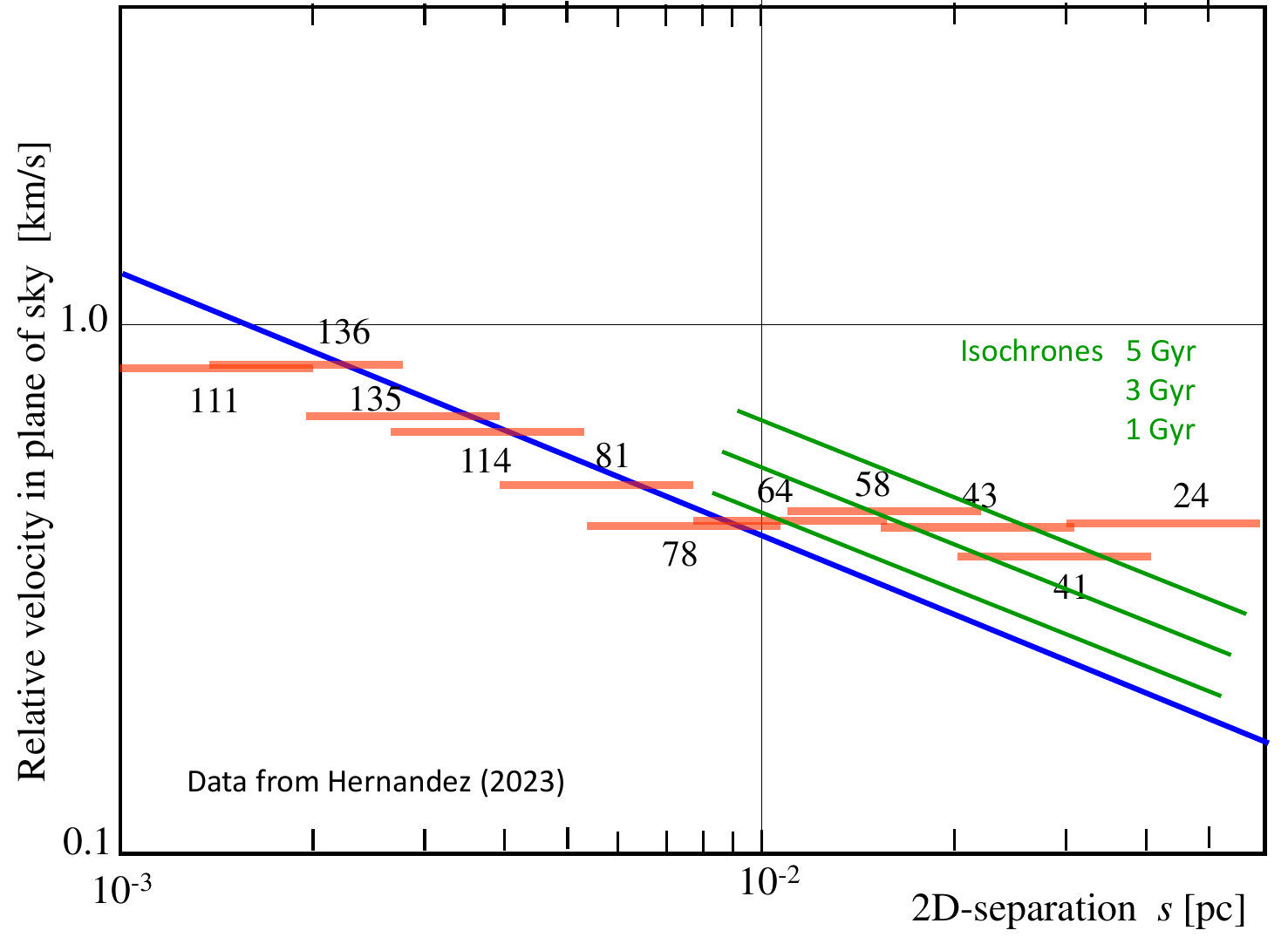}
\caption{Binned root mean square $\langle \Delta V \rangle$ of the relative velocities in the plane of sky as a function of the 2D projected separation $s$ for 450 binary pairs, remaining after the application of most  stringent quality criteria by Hernandez \cite{Hernandez23}. There is a partial overlap of the intervals of binned binary pairs. The black numbers indicate the number of binary pairs in the various overlapping means considered. The green  lines show the  isochrones corresponding to departures from the Newtonian law after 1, 3 and 5 Gyr of evolution under the dynamical acceleration in the scale invariant theory (Adapted from Hernandez \cite{Hernandez23}).}
\label{Hernan23}
\end{figure*}

 The binned  root-mean-square  $\langle \Delta V \rangle$ of the relative velocities between the two components of selected pairs are shown in Fig.~\ref{Hernan23} as a function of the 2D projected separations $s$  between the two components of each binary pair.
In the Figure, this blue line shows the Newtonian relation $\upsilon \sim s^{-1/2}$ resulting of  simulations of 50'000 binary pairs composed of two stars of 1 M$_{\odot}$ mass \citep{Jiang10}. The simulations assume a Newtonian evolution of 10 Gyr under the galactic field in the solar neighborhood, random encounters with field stars, random orientation of the lines of sight and variable eccentricities in agreement with observations. Interestingly enough, this blue line just follows a Keplerian relation $\langle \Delta V \rangle \sim  s^{-1/2}$ up to the Jacobi radius, also called Roche radius. In the present case, it corresponds to the galactocentric distance where the Galactic field would dissolve the binary system. The value of this critical radius is 1.70 pc  in the present case \citep{Jiang10}, thus well beyond the value of about 0.01 pc where the observed relation is turning into a plateau in the figure, as analysed in the next section. Note  for comparison with  Chae \cite{Chae24} that 0.01 pc corresponds to 2.063 kau or also $\log(s) = 3.31$ (see Fig.~\ref{Chae21} for comparison).

If the orbital velocities of the two stars are constant with separation, it is also the case for their differences, a property that allowed building Fig.~\ref{Hernan23}. There, the Keplerian relation, consistent with Newton's Law,  is very well followed for separations up to about  $10^{-2}$ pc. However, for larger separations the mean relative velocities significantly start deviating from the Keplerian relation and tend to an essentially flat constant value  in  Fig.~\ref{Hernan23}. We note that this flat appearance is  favoured by the graphical representation of mean binned values of the velocities, as well as possibly by the objective to perform comparisons with the flat behaviour predicted by MOND. Indeed, the comparison with the isochrones corresponding to the duration of the free evolution gives results similar to those by Chae \cite{Chae24}. The result by \cite{Hernandez23} also  support an evolution in the \emph{dynamical regime}  of the order 5 Gyr, in excellent agreement with the results obtained with the sub-sample of Chae \cite{Chae24} containing the very best quality observations, while the large statistical results of Chae are more in favour of 3 Gyr. The key point is that the different results on the observed velocities of the very wide binary systems are all supporting significant deviations from the standard Newtonian predictions, in a way which is consistent with an evolution dominated by Eq.~(\ref{Nvec6}) for several Gyr.

A possible difficulty has been raised by Loeb \cite{Loeb22}: in  wide binaries the gravitational redshift becomes about similar in amplitude as  the Doppler-Fizeau effect. This might effectively be a problem. However, one has to  consider the differential effects of gravitational redshifts. Following Loeb, we express the velocity shift $\upsilon_{\mathrm{GR}}$ due to gravitation, and we apply the mass-radius relation,
\begin{equation}
\frac{R}{R_{\odot}} \approx  \left(\frac{M}{M_{\odot}} \right)^{0.8},
\end{equation}
for Solar-mass  stars close to the Main Sequence. We then obtain:
\begin{equation}
\upsilon_{\mathrm{GR}} = - \frac{G\, M}{cR}, \quad \quad \mathrm{and \, thus}  \quad
\upsilon_{\mathrm{GR}}  \approx - 0.6  \left(\frac{M}{M_{\odot}}\right)^{0.2}  \; \, \mathrm{km/s}.
\label{ups2}
\end{equation}
The  total binary masses  for the studied systems with large $s$  are confined to a limited interval of 1.2 to 2.0 $M_{\odot}$ \cite{Hernandez23}. The distribution of the mass ratios of binaries in the mass range of solar type stars  was studied by \cite{Duquennoy91}, who found a mean mass ratio $M_2/M_1$ of about 0.5.  To the power of 0.2, this  gives  a value of 0.87 for  the ratio of the gravitational redshifts of the two components. For a velocity of 0.60 km/s for the primary, this corresponds to $\Delta \upsilon_{\mathrm{GR}}$ of about 0.08 km/s, which  is small and cannot be responsible for the observed deviations. Thus, while the remark by \cite{Loeb22} that the gravitational shifts are of the same order as the Doppler-Fizeau shifts is correct, the differences of gravitational redshifts between the two components in a binary pair are significantly smaller than the velocity shifts. 

Most importantly, gravitational redshifts do not introduce any  systematic trend, in the sense that sometimes it is the blue shifted approaching star which has more gravity shift (reducing the $\upsilon$-difference),
and sometimes it is the receding one (increasing the $\upsilon$-difference).  Thus, the velocity differences may be increased or  reduced in a non-systematic way. In the mean of many binary pairs in a given mass interval  the differential effects of the gravitational redshifts of the pairs should statistically cancel each other. These remarks are important for the sample in \cite{Chae24}, while in the case of \cite{Hernandez23}, the relative motions of the two components are mainly derived from  the proper motions in the plane of the sky, orthogonal to the direction  where gravitational effects is acting, as emphasized by this author.

\section{Conclusions} 
\label{concl}

In this work, we consider the very accurate data on wide binary stars obtained from Gaia DR3. We focus in particular on the work by Hernandez \cite{Hernandez23} and Chae \cite{Chae24}, which both suggest too high velocities with respect to standard Newtonian predictions at  separations larger than about 3000 au, or about 0.015 pc. This effect is similar to what is observed in the external parts of galaxy rotation curves, but as stars obsviously do not reside in a dark matter halo, this standard interpretation does not hold anymore.

Instead, we propose a much more simple explanation in the framework of scale invariant theory. This theory rests on one single and simple physical assumption, without any adjustment of any ad'hoc parameter. The direct consequence of scale invariance is that it naturally adds a dynamical acceleration that best manifests itself at low gravities in the modified Newton equation. This is exactly the dynamical regime of wide binary stars. 

We find that the dynamical evolution of very wide binary stars indeed appears to be dominated by the additional dynamical acceleration predicted by the scale invariant theory.  The amplitude of the observed deviations to Newton's gravity supports such a dynamical evolution of loose systems, for several Gyrs. This result is robust against a variety of sub-selections of stars and quality criteria and is not influenced by gravitational redshift effects that we find neglible in front of dynamical evolution. 

This new result of the present series on observational tests of scale invariance is the third positive test after rotation curves of galaxies and galaxy clusters \cite{Maeder24} and strong gravitational lensing \cite{MCourbin24}. It also comes in strong support of previous work on scale invariance explaining the faster-than-expected growth of structure in the Universe \cite{MaedGueor19}, Lunar recession \cite{MGueorguiev21}, and the CMB temperature evolution \cite{Maeder17b}.

 


\acknowledgments

A.M.  expresses his best  thanks to Dr. Vesselin Gueorguiev for constructive interaction since many  years.






\begin{thebibliography}{99}
\bibliographystyle{JHEP}

\bibitem[Auger et al.(2009)]{Auger09} Auger, M.W., Treu, T., Bolton, A.S.  et al., \emph{The SLOAN lens ACS Survey. XI. Colors, lensing and stellar masses of early-type galaxies.},  ApJ, 705, 1099 (2009)



\bibitem{Banik24} Banik, I., Pittordis, C., Sutherland, W. et al., \emph{Strong constraints on the gravitational law from Gaia DR3 wide binaries}, MNRAS, 527, 4573

\bibitem{BKroupa19}
Banik, I., Kroupa, P., \emph{Directly testing gravity with Proxima Centauri}, MNRAS, 487, 1653 (2019)



\bibitem{Belokurov20} Belokurov, V., Penoyre, Z., Oh, S. et al.{\emph{Unresolved stellar companions with Gaia DR2 astrometry}},
MNRAS, 496, 1922 (2020)






\bibitem{Canuto77}
{Canuto}, V., {Adams}, P.~J., {Hsieh}, S.-H., \& {Tsiang}, E.,
\emph{Scale-covariant theory of gravitation and astrophysical applications}, Physical Rev. D, 16, 1643 (1977)


\bibitem{Chae23} Chae, H.-K., \emph{Breakdown of the Newton–Einstein Standard Gravity at Low Acceleration in Internal
Dynamics of Wide Binary Stars}, ApJ, 952, 128 (2023)

\bibitem{Chae24} Chae, H.-K., \emph{Robust Evidence for the Breakdown of Standard Gravity at Low Acceleration from Statistically Pure
Binaries Free of Hidden Companions}, arXiv: 2309.10404 (2024)




\bibitem{Clarke20} Clarke, C.J., \emph{The distribution of relative proper motions of wide binaries in Gaia DR2: MOND or multiplicity?}
 MNRAS, 491, L72 (2020)



\bibitem{Dirac73}
Dirac, P.~A.~M., \emph{Long range forces and broken symmetries}, Proceedings of the Royal Society of London Series A,
333, 403 (1973)

\bibitem{Dirac74}
Dirac, P.~A.~M., \emph{Cosmological Models and the Large Numbers Hypothesis},
 Proceedings of the Royal Society of London Series A, 338, 439
(1974)
\bibitem{Duquennoy91} Duquennoy, A., Mayor, M., \emph{Multiplicity among Solar Type Stars in the Solar Neighbourhood - Part Two - Distribution of the Orbital Elements in an Unbiased Sample},
Astron. Astrophys.248, 485 (1991)




\bibitem{El-Badry21} El-Badry, K., Rix, H.W., Heinz, T.M., \emph{A million binaries from Gaia eDR3: sample selection and validation of Gaia parallax uncertainties},
MNRAS, 506, 2269










\bibitem{GaiaColl21} Gaia Collaboration, Brown, A.G.A., Vallenari, A. et al.,
\emph{Gaia Early Data Release 3. The Gaia Catalogue of Nearby Stars},
Astron.Astrophys. 649, A1 (2021)






\bibitem{Hernandez12} Hernandez, X., Jimenez, M.A., Allen, C., \emph{Wide binaries as a critical test for Gravity theories},
European Physical Journal C, 72, 1884 (2012)

\bibitem{Hernandez19} Hernandez, X., Cortes, R.A.A., Allen, C., 
\& Scarpa, R., \emph{Detailed Solar System dynamics as a probe of the Dark Matter hypothesis}, MNRAS 483, 147 (2019)


\bibitem{Hernandez22} Hernandez, X., Cookson, S., Cortes, R.A.A.,
\emph{Internal kinematics of Gaia eDR3 wide binaries}
MNRAS 502, 2304 (2022)

\bibitem{Hernandez23} Hernandez, X., \emph{Internal kinematics of Gaia DR3 wide binaries: anomalous behaviour in the low acceleration regime}, MNRAS, 525, 1401
(2023)
\bibitem{Hernandez24} Hernandez, X.,  Verteletskyi, V.,  Nasser, L.  and Aguayo-Ortiz, A. 2024,  submitted, arXiv: 2309.10995



\bibitem{Jiang10} Jiang, Y.-F.,
Tremaine, S., \emph{The evolution of wide binary stars}, MNRAS, 401, 977 (2010)

 
 




 
 
 




 



\bibitem{Loeb22} Loeb, A., \emph{Gravitational Redshift for Wide Binaries in Gaia eDR3}, Res. Note AAS, 6, 55 (2022)



\bibitem{Maeder78} Maeder, A., \emph{Metrical connection in space-time, Newton's and Hubble's laws}, Astron. Astrophys. 65, 337 (1978)


\bibitem{Maeder17a} Maeder, A.,
\emph{An Alternative to the LambdaCDM Model: The Case of Scale Invariance}
Astrophys. J., 834, 194 (2017a)


\bibitem{Maeder17b} Maeder, A., \emph{Scale-invariant Cosmology and CMB Temperatures as a Function of Redshifts} Astrophys. J., 847, 65 (2017b)

\bibitem{Maeder17c} Maeder, A., \emph{
Dynamical Effects of the Scale Invariance of the Empty Space: The Fall of Dark Matter?}
Astrophys. J., 849, 158 (2017c)




\bibitem{Maeder23} Maeder, A., \emph{MOND as a peculiar case of the SIV theory},  MNRAS, 520, 1447  (2023)

\bibitem{Maeder24} Maeder, A. 2024, \emph{Observational tests in scale invariance I: galaxy clusters and rotation of galaxies},  submitted   (2024)    

\bibitem{MBouvier79} Maeder, A., Bouvier, P., \emph{ Scale invariance, metrical connection and the motions of astronomical bodies},
 Astron. Astrophys., 73, 82 (1979)

\bibitem{MCourbin24} Maeder, A., Courbin, F.,  \emph{Observational tests in scale invariance II: gravitational lensing}, submitted (2024) \%lensing


\bibitem{MaedGueor19} Maeder, A.; Gueorguiev, V.G., \emph{The growth of the density fluctuations in the scale-invariant vacuum theory}, Physics of the Dark Universe, 25, 100315 (2019)

\bibitem{MGueorguiev21} Maeder, A.; Gueorguiev, V.G., \emph{On the relation of the lunar recession and the length-of-the-day}, Astrophys Space Science, 366, 101 (2021)







\bibitem{MaedGueor23}
Maeder, A.; Gueorguiev, V.G., \emph{Action Principle for Scale Invariance and Applications (Part I)}, Symmetry 2023, 15, 1966; arXiv 2310.16913 (2023)








\bibitem{Milgrom83} Milgrom, M., \emph{A modification of the newtonian dynamics : implications for galaxy systems}, Astrophys. J., 270, 365 (1983)


\bibitem{Milgrom09} Milgrom, M., \emph{The Mond Limit from Spacetime Scale Invariance}, Astrophys. J., 698, 1630 (2009)

















\bibitem{Nestor Shachar23} Nestor Shachar, A., Price, S.H., Forster Schreiber, N. M. et
al., \emph{RC100: Rotation Curves of 100 Massive Star-forming Galaxies at z = 0.6-2.5 Reveal 
Little Dark Matter on Galactic Scales}, ApJ 944, 78 (2023)





\bibitem{Pittordis18}
Pittordis, C., Sutherland, W., \emph{Testing modified-gravity theories via wide binaries and GAIA},  MNRAS, 480, 1778  (2018)

\bibitem{Pittordis23}
Pittordis, C., Sutherland, W., \emph{WIDE BINARIES FROM GAIA EDR3: PREFERENCE FOR GR OVER MOND ?}, Open Journ. Astrophys. 6, 4 (2023)








\bibitem{Scarpa17} Scarpa, R., Ottolina, R., Falomo, R. et al., \emph{Dynamics of wide binary stars: A case study for testing Newtonian dynamics in the low acceleration regime}, 
IJMPD, 26, 1750067 (2017)

  

  

  
  
  




 
\bibitem{Dokkum23} van Dokkum, P., Brammer, G., Wang, B. et al.,
 \emph{A massive compact quiescent galaxy at z=2 with a complete Einstein ring in JWST imaging}, Nature Astronomy, 8, 119 (2023)


 
 

  
  
 





\end{thebibliography}
\end{document}